\documentclass[preprint,12pt]{elsarticle}
\usepackage{url}
\usepackage{booktabs}
\usepackage[table]{xcolor}
\usepackage{array}
\usepackage{amssymb}
\usepackage{rotating}
\newcommand{\cmark}{\checkmark}
\newcommand{\xmark}{\(\times\)}
\newcommand{\pmark}{\(\triangle\)}

\journal{International Journal of Information Technology \& Decision Making}

\begin{document}

\begin{frontmatter}

\title{A Tumour-Aware Multimodal Framework for Explainable Prediction of IDH Mutation in Gliomas from Multi-Sequence MRI\tnoteref{t1}}
\tnotetext[t1]{Preprint of an article submitted for consideration in International Journal of Information Technology \& Decision Making \copyright\ 2026 copyright World Scientific Publishing Company, \url{https://www.worldscientific.com/worldscinet/ijitdm}}

\author[inst1,inst2]{Tahsien Al-Quraishi\corref{cor1}}
\ead{tahsien.alquraishi@gmail.com}
\cortext[cor1]{Corresponding author}
\affiliation[inst1]{organization={School of Business, Melbourne Institute of Technology},
  city={Melbourne}, postcode={3000}, state={Victoria}, country={Australia}}
\affiliation[inst2]{organization={Al-Ayen Iraqi University}, city={Nasiriyah}, country={Iraq}}

\author[inst3]{Waseem Alwan Zaboon}
\affiliation[inst3]{organization={Mechanical Engineering Department, College of Engineering, Wasit University},
  city={Al Kut}, country={Iraq}}

\author[inst4]{O.S. Albahri}
\affiliation[inst4]{organization={Federation University Australia}, city={Ballarat}, state={Victoria}, country={Australia}}

\author[inst5]{Raza Nowrozy}
\affiliation[inst5]{organization={College of Art, Business, Law, Education and IT (CoABLEIT), Victoria University},
  city={Melbourne}, postcode={3000}, state={Victoria}, country={Australia}}

\begin{abstract}

Accurate prediction of isocitrate dehydrogenase (IDH) mutation status is essential for glioma diagnosis, prognosis, and treatment planning. Although recent deep learning approaches have demonstrated promising performance, many rely on whole-brain feature extraction and provide limited interpretability. This study proposes a tumour-aware multimodal framework for explainable, non-invasive prediction of IDH mutation status from multiparametric magnetic resonance imaging (MRI). The proposed framework integrates four MRI sequences (T1, T1CE, T2, and FLAIR), tumour subregion segmentation, quantitative radiomic descriptors, and clinical information within a unified deep learning architecture. An ablation study was conducted to evaluate the contribution of tumour-aware learning, radiomic features, and clinical variables. Model performance was assessed using multiple complementary evaluation metrics, including accuracy, precision, recall, F1-score, balanced accuracy, receiver operating characteristic analysis, confusion matrix evaluation, and Grad-CAM visualisation. Experimental results demonstrated that the proposed multimodal framework consistently outperformed the baseline MRI-only model, with the Clinical Fusion model achieving the strongest overall predictive performance. The integration of tumour-aware learning, radiomic descriptors, and clinical information progressively improved classification performance, while Grad-CAM visualisations indicated that the model predominantly focused on biologically relevant tumour regions. These findings suggest that tumour-aware multimodal feature integration enables accurate, interpretable, and non-invasive molecular profiling of gliomas and provides a promising foundation for future clinical decision support in precision neuro-oncology.

\end{abstract}


\end{frontmatter}

\section{Introduction}

Gliomas are the most common primary malignant tumours of the central nervous system and are characterised by considerable molecular and clinical heterogeneity \cite{cui2025towards}. Their prognosis, treatment response, and overall survival vary substantially according to tumour grade and underlying genetic alterations. Consequently, recent editions of the World Health Organization (WHO) Classification of Central Nervous System Tumours have incorporated molecular biomarkers alongside histopathological findings, reflecting the increasing importance of molecular profiling in precision neuro-oncology \cite{louis20212021}. Among the recognised molecular biomarkers, isocitrate dehydrogenase (IDH) mutation status has emerged as one of the most important diagnostic and prognostic indicators for diffuse gliomas. Patients with IDH-mutated gliomas generally experience slower disease progression, improved treatment response, and longer overall survival than those with IDH-wildtype tumours \cite{fang2026isocitrate}. Accurate determination of IDH mutation status, therefore, plays an essential role in tumour classification, prognostic assessment, treatment planning, and personalised patient management \cite{li2022noninvasive}. At present, definitive identification of IDH status relies on histopathological examination and molecular testing following surgical biopsy or tumour resection \cite{guo2023histological}.

Although these approaches provide reliable molecular information, they are invasive, carry procedural risks, and are unsuitable for repeated assessment in routine disease monitoring \cite{song2025advancements}. Consequently, there is growing interest in developing accurate, non-invasive imaging-based methods capable of predicting molecular characteristics before surgical intervention and supporting preoperative clinical decision-making \cite{christodoulou2025artificial}. Multi-parametric magnetic resonance imaging (MRI) is routinely used for the diagnosis and follow-up of patients with gliomas because it provides comprehensive anatomical and physiological information without exposing patients to ionising radiation \cite{wang2022multi}. Recent advances in artificial intelligence, particularly deep learning, have enabled automated extraction of complex imaging features from MRI for tumour classification and molecular prediction \cite{khalighi2024artificial}. Numerous studies have reported encouraging results using convolutional neural networks, radiomics, and multimodal learning strategies for predicting IDH mutation status \cite{zhong2022predicting}. However, many existing approaches rely on whole-image feature extraction, allowing irrelevant background tissue to influence the learning process and limiting the ability to capture tumor-specific characteristics \cite{reig2020machine}. 

In addition, relatively few studies combine tumour-aware learning, quantitative radiomic descriptors, clinical variables, and explainable artificial intelligence (AI) within a single unified framework \cite{singh2026saasu}. Tumour-aware learning provides an opportunity to overcome these limitations by directing feature extraction towards biologically relevant tumour regions identified through segmentation \cite{astley2022deep}. Radiomic descriptors provide complementary quantitative information describing tumour morphology, intensity, texture, and heterogeneity, while clinical variables such as patient age, sex, and tumour grade contribute additional prognostic information \cite{bera2022predicting}. Furthermore, explainable AI techniques improve model transparency by highlighting the image regions that contribute to each prediction, thereby increasing confidence in AI-assisted clinical decision-making \cite{abbas2025explainable}.

To address these challenges, this study proposes a tumour-aware multimodal framework for explainable prediction of IDH mutation status from multi-parametric MRI. The proposed framework integrates four MRI sequences (T1, T1CE, T2, and FLAIR), tumour subregion segmentation, quantitative radiomic descriptors, and clinical information within a unified deep learning architecture. The framework is designed to improve predictive performance while maintaining model interpretability through saliency-based explainability \cite{brima2024saliency}. Comprehensive experimental evaluation, including ablation studies, receiver operating characteristic analysis, confusion matrix analysis, and explainability assessment, was conducted to assess the contribution of each component to non-invasive molecular prediction.

\subsection{Research Gap, Novelty, and Contributions}

Despite the encouraging progress achieved by recent deep learning methods for MRI-based IDH mutation prediction, several important challenges remain \cite{bangalore2025bridging}. Many existing studies rely primarily on global image features and do not explicitly guide feature learning towards tumour-specific regions \cite{sasikala2026exu}. In addition, relatively few approaches integrate tumour-aware learning, radiomic descriptors, clinical information, and explainable AI within a unified prediction framework. These limitations reduce model interpretability and may limit the ability to capture the complementary information required for accurate molecular characterisation \cite{alhamrani2025machine}. The aim of this study is to develop an explainable tumour-aware multimodal framework for accurate preoperative prediction of IDH mutation status from multi-parametric MRI. By combining tumour subregion segmentation, radiomic descriptors, and clinical variables within a unified deep learning framework, the proposed approach seeks to improve predictive accuracy while providing interpretable predictions suitable for clinical decision support.

The primary contribution of this study is the unified integration of tumouraware learning, quantitative radiomic analysis, clinical information, and explainable AI within a single framework for non-invasive prediction of IDH mutation status. Although these individual components have been investigated in previous studies, relatively few approaches have integrated them within a unified and interpretable framework for MRI-based IDH mutation prediction~\cite{vanguri2022multimodal}. Unlike conventional MRIbased deep learning models that analyse the entire brain volume, the proposed framework explicitly guides feature extraction towards biologically relevant tumour regions while incorporating complementary quantitative radiomic descriptors and patient-specific clinical variables. Furthermore, gradient-based saliency analysis provides visual explanations of the model’s predictions, improving model transparency and supporting future clinical translation~\cite{brima2024saliency}. 3

The main contributions of this study are summarised as follows:

\begin{itemize}

\item A tumour-aware multimodal deep learning framework is proposed for preoperative prediction of IDH mutation status by integrating multi-parametric MRI, tumour subregion segmentation, radiomic descriptors, and clinical variables.

\item An ablation study is performed to quantify the individual contributions of tumour-aware learning, radiomic features, and clinical information to prediction performance.

\item An explainable AI framework based on gradient-based saliency analysis is introduced to improve model interpretability and verify tumour-focused feature learning.

\item Experimental results demonstrate that multimodal feature integration substantially outperforms conventional MRI-only approaches for accurate and reliable preoperative IDH mutation prediction.

\end{itemize}

The remainder of this paper is organised as follows. Section~2 reviews the related literature on MRI-based IDH mutation prediction, tumour-aware learning, radiomics, and explainable AI. Section~3 describes the proposed methodology, including the study dataset, image preprocessing, tumour segmentation, radiomic feature extraction, multimodal framework, and experimental setup. Section~4 presents the experimental results and performance evaluation. Section~5 discusses the findings, compares the proposed framework with existing studies, and outlines its clinical relevance, limitations, and future research directions. Finally, Section~6 concludes the paper and summarises the main contributions of this study.

\section{Related Work}

Recent advances in AI have transformed non-invasive molecular profiling of gliomas using multiparametric MRI \cite{liang2024artificial,agosti2026mri}. Deep learning, radiomics, multimodal learning, and XAI have demonstrated considerable potential for predicting clinically important biomarkers, particularly IDH mutation status \cite{christodoulou2025artificial}. Accurate preoperative prediction of IDH mutations can support tumour classification, prognosis, and personalised treatment planning while reducing dependence on invasive biopsy procedures \cite{khalighi2024artificial}. Despite substantial progress, important challenges remain regarding tumour localisation, multimodal feature integration, model interpretability, and clinical generalisability \cite{boehm2022harnessing}. This section critically reviews recent developments in MRI-based IDH mutation prediction and highlights the methodological gaps that motivate the proposed tumour-aware multimodal framework.

\subsection{Deep Learning for MRI-Based IDH Mutation Prediction}

Deep learning has become the dominant approach for MRI-based prediction of IDH mutation status owing to its ability to automatically learn hierarchical imaging representations from large-scale datasets \cite{christodoulou2025artificial}. Early convolutional neural network (CNN)-based models demonstrated that multiparametric MRI contains sufficient imaging biomarkers for reliable molecular classification, eliminating the need for handcrafted feature engineering \cite{rundo2024image}. Choi et al. \cite{choi2021fully} proposed a fully automated hybrid framework integrating tumour segmentation, CNN-based classification, and radiomic analysis, demonstrating strong predictive performance across both internal and external validation cohorts. Although their study established the feasibility of automated molecular prediction, explainability remained limited and feature learning was not explicitly constrained to biologically relevant tumour regions. Subsequent studies extended deep learning beyond binary IDH mutation prediction towards comprehensive molecular glioma classification such as Cluceru et al. \cite{cluceru2022improving} demonstrated that deep neural networks could classify multiple WHO 2021 molecular glioma subtypes using MRI-derived imaging features, highlighting the capability of deep learning to capture complex molecular heterogeneity. Nevertheless, these approaches primarily focused on comprehensive glioma classification rather than optimising models specifically for IDH mutation prediction. Furthermore, clinical variables, tumour-aware learning, and explainability were generally investigated independently rather than integrated within a unified prediction framework \cite{choi2021fully,cluceru2022improving,christodoulou2025artificial}.

More recently, transformer-based architectures have attracted increasing attention because of their ability to capture long-range spatial dependencies and complex contextual relationships within MRI volumes \cite{zhu2025transformer}. Niu et al. \cite{niu2025mri} combined transformer-based imaging representations with radiomic descriptors for molecular glioma prediction, reporting improved discriminative performance compared with conventional CNN architectures. Nevertheless, transformer models generally require substantially larger training datasets, exhibit greater computational complexity, and often remain difficult to interpret clinically \cite{abibullaev2023deep}. Their deployment in routine clinical practice therefore remains challenging \cite{christodoulou2025artificial}. Overall, existing deep learning approaches have achieved encouraging predictive performance but exhibit several important limitations \cite{barbierato2024challenges}. Most frameworks perform feature extraction from the entire brain volume, allowing non-tumour tissue to contribute to feature learning and potentially reducing biological specificity \cite{siddiqui2025hybrid}. Moreover, relatively few studies integrate tumour-aware feature learning, quantitative radiomic descriptors, clinical variables, and explainable AI within a unified prediction framework \cite{choi2021fully,niu2025mri,christodoulou2025artificial,mojtahedi2025deep}. These limitations provide the primary motivation for the proposed tumour-aware multimodal approach.

\subsection{Radiomics, Multimodal Learning, and Explainable AI}

Radiomics complements deep learning by extracting quantitative imaging biomarkers that describe tumour morphology, intensity, texture, and spatial heterogeneity \cite{avanzo2020machine}. Unlike automatically learned deep features, handcrafted radiomic descriptors provide interpretable quantitative measurements that can be directly associated with tumour phenotype and biological behaviour \cite{bera2022predicting}. Consequently, radiomics has become an important component of imaging-based molecular profiling \cite{rathore2018radiomic}. Several studies have demonstrated that combining radiomics with deep learning improves IDH mutation prediction \cite{hosseini2023mri,li2017deep,niu2025mri}. Choi et al. \cite{choi2021fully} reported that integrating handcrafted radiomic descriptors with CNN-derived features enhanced molecular classification performance by capturing complementary tumour characteristics. Similarly, Truong et al. \cite{truong2024two}proposed a multicontrast MRI radiomics framework for preoperative IDH mutation prediction, demonstrating that quantitative tumour descriptors extracted from multiple MRI sequences significantly improved predictive performance compared with single-modality analysis.

Generalisability remains one of the principal challenges for radiomics-based prediction models. Lost et al. \cite{lost2024enhancing} addressed this issue through external validation of a machine learning radiomics model using independent patient cohorts. Although their findings demonstrated encouraging robustness, the framework relied primarily on handcrafted radiomic features and did not incorporate deep imaging representations, tumour-aware learning, or XAI. Consequently, the complementary advantages of multimodal feature integration were not fully exploited. Recent systematic reviews and meta-analyses further support the diagnostic potential of radiomics and deep learning for MRI-based molecular prediction. Chen et al. concluded that radiomics models consistently achieve favourable diagnostic accuracy for predicting IDH mutation status but identified substantial methodological heterogeneity, inconsistent validation strategies, and limited reproducibility across studies. Likewise, Farahani et al. \cite{farahani2025diagnostic} highlighted the growing clinical potential of multimodal deep learning while emphasising the need for standardised preprocessing pipelines, external validation, and improved model interpretability before routine clinical implementation.

Model explainability has emerged as another critical requirement for clinical adoption. Although many recent studies report excellent predictive performance, relatively few provide convincing evidence that their predictions are driven by biologically meaningful tumour regions. Post hoc explainability techniques, including Grad-CAM and saliency analysis, have therefore become increasingly important for verifying model behaviour and improving clinician confidence in AI-assisted decision-making. Nevertheless, explainability remains insufficiently integrated into many existing molecular prediction frameworks \cite{selvaraju2020grad}.

\begin{table*}[!t]
\centering
\scriptsize
\setlength{\tabcolsep}{3pt}
\renewcommand{\arraystretch}{1.2}

\caption{Comparison of MRI-based IDH mutation prediction methods}
\label{tab:comparison}

\begin{tabular}{p{2.8cm}c*{6}{>{\centering\arraybackslash}p{0.55cm}}p{4.6cm}}
\toprule
\textbf{Study} &
\textbf{Year} &
\textbf{DL} &
\textbf{TA} &
\textbf{Rad} &
\textbf{Clin} &
\textbf{XAI} &
\textbf{EV} &
\textbf{Key Innovation} \\
\midrule

Choi et al.~\cite{choi2021fully}
& 2021
& \cmark
& \cmark
& \cmark
& \xmark
& \xmark
& \cmark
& Fully automated CNN--radiomics framework \\

Cluceru et al.~\cite{cluceru2022improving}
& 2022
& \cmark
& \xmark
& \xmark
& \xmark
& \xmark
& \xmark
& Deep learning for WHO 2021 molecular glioma classification \\

Truong et al.~\cite{truong2024two}
& 2024
& \pmark
& \cmark
& \cmark
& \xmark
& \xmark
& \cmark
& Two-stage multicontrast MRI radiomics \\

Lost et al.~\cite{lost2024enhancing}
& 2024
& \xmark
& \xmark
& \cmark
& \xmark
& \xmark
& \cmark
& Externally validated radiomics model \\

Siddiqui et al.~\cite{siddiqui2025hybrid}
& 2025
& \cmark
& \xmark
& \pmark
& \xmark
& \xmark
& \xmark
& Hybrid whole-brain MRI framework \\

Niu et al.~\cite{niu2025mri}
& 2025
& \cmark
& \xmark
& \cmark
& \xmark
& \xmark
& \cmark
& Transformer--radiomics fusion \\

\midrule

\rowcolor{gray!20}
\textbf{This Study}
&
\textbf{2026}
&
\textbf{\cmark}
&
\textbf{\cmark}
&
\textbf{\cmark}
&
\textbf{\cmark}
&
\textbf{\cmark}
&
\textbf{\xmark}
&
\textbf{Tumour-aware multimodal fusion integrating deep learning, radiomics, clinical variables, and Grad-CAM explainability} \\

\bottomrule
\end{tabular}

\end{table*}

Table~\ref{tab:comparison} presents a methodological comparison of representative MRI-based studies published between 2021 and 2026 for preoperative prediction of IDH mutation status. Rather than comparing predictive performance alone, the table evaluates the key methodological components adopted by existing approaches, including deep learning, tumour-aware feature extraction, radiomics, clinical variables, explainable artificial intelligence (XAI), external validation, and multimodal feature fusion. Several important observations emerge from this comparison. First, deep learning has become the dominant modelling paradigm, with recent studies employing convolutional neural networks, hybrid CNN--radiomics frameworks, and transformer-based architectures to improve molecular prediction performance \cite{christodoulou2025artificial}. Despite these advances, only a limited number of studies explicitly incorporate tumour-aware learning to constrain feature extraction to biologically relevant tumour regions. Most existing approaches continue to rely on whole-brain image representations, allowing non-tumour tissue to influence feature learning and potentially reducing biological specificity.

Second, radiomics has become one of the most widely adopted complementary techniques for characterising tumour morphology, intensity, texture, and intratumoural heterogeneity \cite{avanzo2020machine}. Several studies have demonstrated that combining radiomic descriptors with deep learning improves predictive performance compared with single-modality approaches. However, simultaneous integration of radiomic features, clinical variables, tumour-aware learning, and explainable AI remains uncommon. Consequently, the complementary prognostic value of imaging, radiomic, and clinical information has not been fully exploited. Third, model interpretability remains a major barrier to clinical translation \cite{holzinger2020measuring}. Although post hoc explainability techniques, such as Grad-CAM and saliency visualisation, have become increasingly popular, they are incorporated in relatively few molecular prediction frameworks, and quantitative assessment of explanation quality is rarely reported. Similarly, although external validation has become increasingly common, no existing framework simultaneously combines tumour-aware feature extraction, deep learning, radiomics, clinical variables, explainable AI, and robust external validation within a single unified prediction pipeline.

Overall, the comparison demonstrates that most existing studies optimise individual methodological components rather than adopting a comprehensive multimodal strategy. Motivated by these observations, the proposed framework integrates tumour-aware feature extraction, multi-sequence MRI, quantitative radiomic descriptors, patient-specific clinical variables, and Grad-CAM-based explainability within a unified deep learning architecture. Compared with previous studies, the proposed framework provides one of the most comprehensive integrations of complementary imaging, radiomic, and clinical information for non-invasive prediction of IDH mutation status while maintaining model interpretability. Although external validation remains future work, the proposed framework establishes a robust and clinically interpretable foundation for preoperative molecular profiling of gliomas.

\section{Materials and Methods}

This section describes the methodology used to develop and evaluate the proposed tumour-aware multimodal framework for non-invasive prediction of IDH mutations. It includes details of the study dataset, image preprocessing, tumour segmentation, radiomic feature extraction, model architecture, multimodal feature integration, experimental design, and performance evaluation. The overall workflow was designed to investigate the contribution of tumour-aware learning, radiomic descriptors, and clinical information to accurate and interpretable prediction of IDH mutation status from multiparametric MRI.

\subsection{Patient Cohort and Data Collection}

The study cohort was obtained from the RSNA--ASNR--MICCAI BraTS 2021 glioma dataset, publicly available through The Cancer Imaging Archive (TCIA) \cite{tcia_brats2021}. The original dataset comprised 625 patients with corresponding clinical information, multiparametric MRI examinations, and expert-annotated tumour segmentation masks. Three patients with missing IDH mutation status were excluded, resulting in a final cohort of 622 patients. Table~\ref{tab:dataset_characteristics} summarises the demographic, clinical, and imaging characteristics of the study population. The cohort included a diverse range of glioma subtypes, including glioblastoma, astrocytoma, oligodendroglioma, and mixed glioma variants, providing a heterogeneous and clinically representative dataset. Each patient had four MRI modalities available, namely T1-weighted (T1), contrast-enhanced T1-weighted (T1CE), T2-weighted (T2), and fluid-attenuated inversion recovery (FLAIR) images. The corresponding tumour segmentation masks were used for tumour-aware feature extraction and quantitative radiomics analysis. In addition, clinical variables, including patient age, sex, and tumour grade, were integrated with imaging-derived features within the proposed multimodal framework for non-invasive prediction of IDH mutation status.

\begin{table}[htbp]
\caption{Characteristics of the study cohort}
\label{tab:dataset_characteristics}
\centering
\begin{tabular}{lc}
\hline
Characteristic & Value \\
\hline
Original cohort size & 625 \\
Excluded (missing IDH status) & 3 \\
Final cohort size & 622 \\
Age (years) & 55.0 $\pm$ 15.5 (18--85) \\
Male & 369 (59.3\%) \\
Female & 253 (40.7\%) \\
IDH-wildtype & 446 (71.7\%) \\
IDH-mutated & 176 (28.3\%) \\
Grade II & 101 (16.2\%) \\
Grade III & 97 (15.6\%) \\
Grade IV & 417 (67.0\%) \\
Missing grade information & 7 (1.1\%) \\
Glioblastoma & 384 (61.7\%) \\
Complete MRI availability & 622 (100\%) \\
Available tumour masks & 622 (100\%) \\
MRI modalities & T1, T1CE, T2, FLAIR \\
\hline
\end{tabular}
\end{table}

Following the exclusion of three patients with missing IDH mutation status, the final dataset comprised 622 patients with complete multiparametric MRI examinations and tumour segmentation masks. The cohort exhibited a moderate class imbalance, with IDH-wildtype cases representing 71.7\% of the dataset. Grade IV gliomas constituted the largest tumour grade category (67.0\%), while glioblastoma was the predominant tumour subtype (61.7\%). The availability of complete imaging, clinical, and molecular information provided a comprehensive foundation for developing and evaluating the proposed tumour-aware multimodal framework for IDH mutation prediction. 

\subsection{MRI Acquisition and Preprocessing}
Four MRI modalities were available for each patient, including T1-weighted (T1), contrast-enhanced T1-weighted (T1CE), T2-weighted (T2), and fluid-attenuated inversion recovery (FLAIR) images. The MRI examinations were acquired using scanners from multiple vendors, including Siemens and GE Healthcare, across a range of scanner models and magnetic field strengths, thereby reflecting the heterogeneity encountered in routine clinical practice. This diversity of imaging hardware facilitated evaluation of the proposed framework under realistic acquisition conditions. Prior to model development, the MRI data were verified to ensure the availability of all required imaging modalities and corresponding tumour segmentation masks. Each patient's imaging data included expert-annotated tumour segmentation masks, which were used for tumour-aware feature extraction and quantitative radiomics analysis. The four MRI modalities were co-registered and resampled to a common spatial resolution, followed by intensity normalisation to reduce inter-subject intensity variability and improve model generalisation. Subsequently, the preprocessed MRI volumes and tumour masks were resized to a fixed spatial dimension and stored in NumPy format to facilitate efficient deep learning training. These standardised data were then used for tumour-aware feature extraction, multimodal feature fusion, and prediction of IDH mutation status.

\begin{figure}[htbp]
\centering
\includegraphics[width=0.65\textwidth]{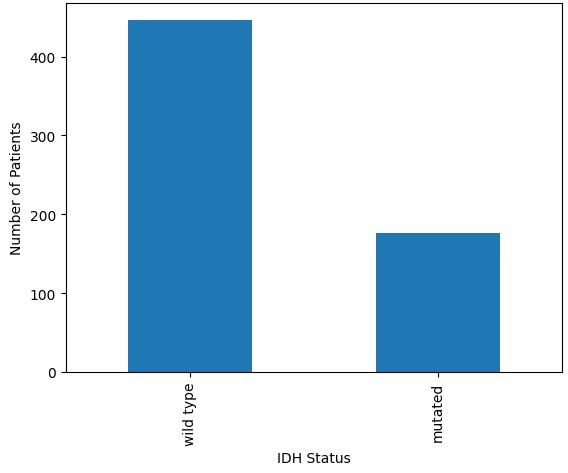}
\caption{Distribution of IDH mutation status in the study cohort}
\label{fig:idh_distribution}
\end{figure}

Figure~\ref{fig:idh_distribution} illustrates the distribution of IDH mutation status within the study cohort. Following the exclusion of three patients with missing IDH information, the final dataset comprised 622 patients, including 446 (71.7\%) IDH-wildtype and 176 (28.3\%) IDH-mutated cases. The observed class imbalance reflects the prevalence of IDH-wildtype gliomas in clinical practice and highlights the importance of adopting appropriate strategies during model development to minimise classification bias.

\begin{figure}[htbp]
\centering
\includegraphics[width=0.70\textwidth]{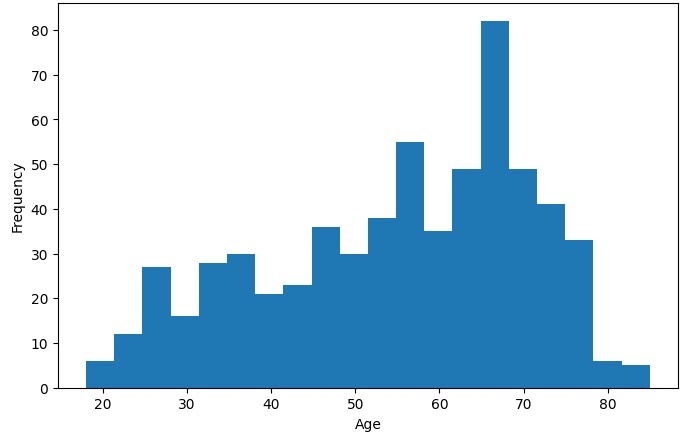}
\caption{Age distribution of patients included in the study cohort.}
\label{fig:age_distribution}
\end{figure}

Figure~\ref{fig:age_distribution} illustrates the age distribution of the study cohort. Patient ages ranged from 18 to 85 years, with a mean age of 55.0 $\pm$ 15.5 years. The distribution demonstrates broad representation across adult glioma patients and supports the development of a robust predictive framework across a clinically diverse population.

\begin{figure}[htbp]
\centering
\includegraphics[width=0.65\textwidth]{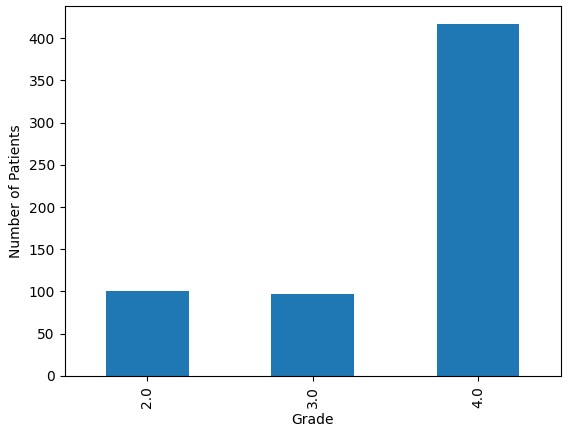}
\caption{Distribution of tumour grades in the study cohort.}
\label{fig:grade_distribution}
\end{figure}

Figure~\ref{fig:grade_distribution} presents the distribution of tumour grades within the study cohort. Grade IV gliomas constituted the largest proportion of cases (67.0\%), whereas Grade II and Grade III gliomas accounted for 16.2\% and 15.6\% of the cohort, respectively. This distribution reflects the predominance of high-grade gliomas in the dataset and provides a representative clinical population for evaluating the proposed framework.

\begin{figure}[htbp]
\centering
\includegraphics[width=0.65\textwidth]{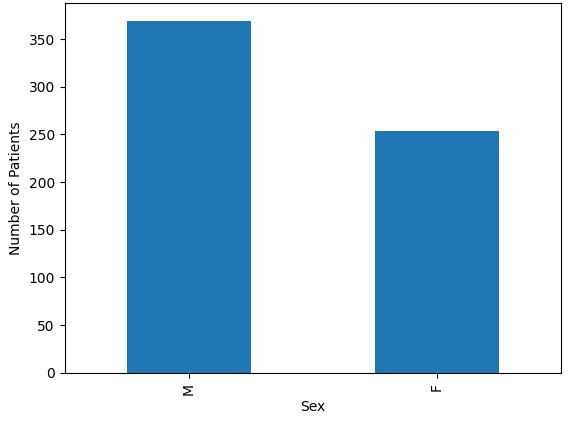}
\caption{Sex distribution of patients included in the study cohort.}
\label{fig:sex_distribution}
\end{figure}

Figure~\ref{fig:sex_distribution} illustrates the sex distribution of the study cohort. The dataset comprised 369 (59.3\%) male and 253 (40.7\%) female patients, indicating a moderate male predominance that is consistent with the reported epidemiology of gliomas. The inclusion of both sexes enhances the representativeness of the study cohort for evaluating the proposed multimodal framework.

\begin{figure}[htbp]
\centering
\includegraphics[width=0.75\textwidth]{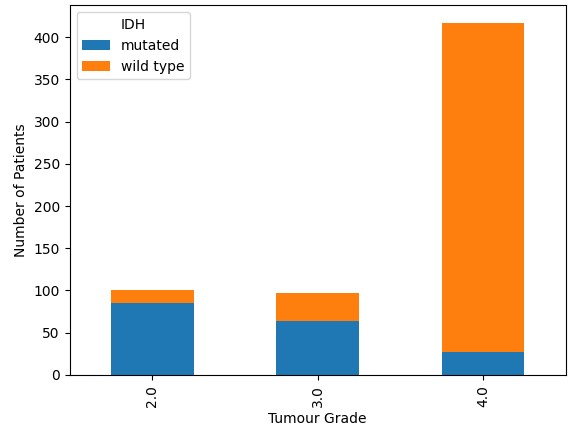}
\caption{Distribution of IDH mutation status across tumour grades.}
\label{fig:idh_grade_counts}
\end{figure}

Figure~\ref{fig:idh_grade_counts} presents the distribution of IDH mutation status across tumour grades within the study cohort. IDH-mutated tumours were predominantly observed in Grade II and Grade III gliomas, whereas Grade IV gliomas were largely associated with the IDH-wildtype phenotype. These findings are consistent with established clinical evidence indicating that IDH mutations occur more frequently in lower-grade gliomas and are considerably less common in aggressive Grade IV gliomas \cite{mcnamara20222021}.

\begin{figure}[htbp]
\centering
\includegraphics[width=0.75\textwidth]{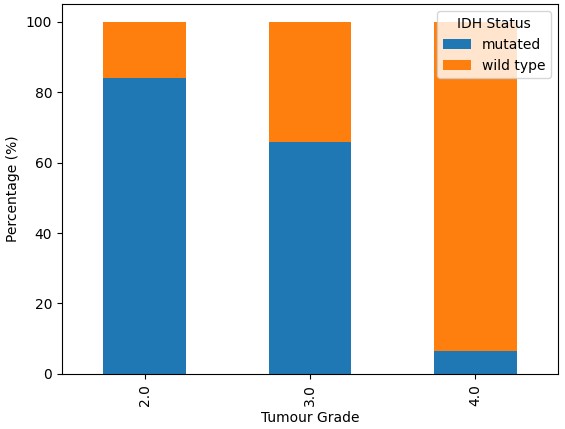}
\caption{Percentage of IDH mutation status within each tumour grade.}
\label{fig:idh_grade_percentage}
\end{figure}

Figure~\ref{fig:idh_grade_percentage} illustrates the percentage of IDH mutation status within each tumour grade. IDH-mutated tumours constituted approximately 84\% of Grade II gliomas and 66\% of Grade III gliomas, whereas Grade IV gliomas were predominantly IDH-wildtype, with only a small proportion of cases harbouring IDH mutations. These findings are consistent with established clinical evidence indicating that IDH mutations are more frequently associated with lower-grade gliomas and are less common in aggressive Grade IV tumours \cite{mcnamara20222021}. The observed relationship further supports the inclusion of tumour grade as a clinically relevant feature within the proposed multimodal framework. The multiparametric MRI dataset comprised four imaging modalities: T1-weighted (T1), contrast-enhanced T1-weighted (T1CE), T2-weighted (T2), and fluid-attenuated inversion recovery (FLAIR). The MRI volumes were stored in NIfTI format with a typical spatial dimension of 240 $\times$ 240 $\times$ 155 voxels. Corresponding expert-annotated tumour segmentation masks were available for all patients. Prior to model development, all MRI volumes and tumour masks underwent intensity normalisation, spatial standardisation, resampling, and resizing to a fixed input resolution to ensure consistency across patients. The preprocessed MRI data were subsequently used for tumour-aware feature extraction, multimodal fusion, and IDH mutation prediction.

\subsection{Tumour Segmentation and Region Extraction}

Tumour segmentation was performed to identify biologically relevant regions for tumour-aware feature extraction and radiomics analysis. Rather than analysing the entire brain volume, the proposed framework focuses on segmented tumour subregions to minimise the influence of surrounding healthy tissue and to extract imaging characteristics more closely associated with the molecular phenotype of gliomas. The tumour masks were subsequently incorporated into both the deep learning and radiomics pipelines to guide feature extraction from clinically meaningful regions of interest \cite{choi2021fully}.

\begin{figure}[htbp]
\centering
\includegraphics[width=0.55\textwidth]{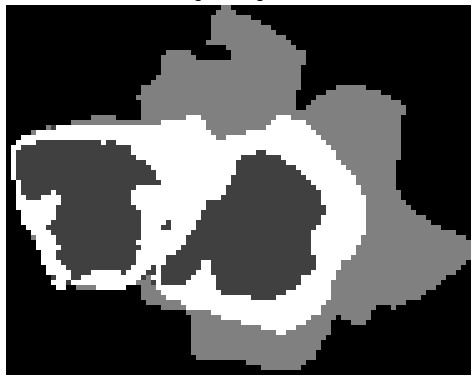}
\caption{Representative tumour segmentation mask}
\label{fig:tumour_mask}
\end{figure}

\begin{figure}[htbp]
\centering
\includegraphics[width=0.90\textwidth]{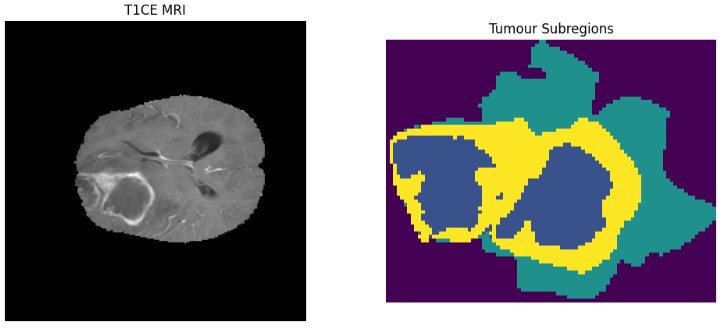}
\caption{Representative T1CE MRI and tumour segmentation mask}
\label{fig:tumour_subregions}
\end{figure}
\newpage 

Figure~\ref{fig:tumour_mask} illustrates a representative tumour segmentation mask containing the tumour subregions used throughout the proposed framework. Figure~\ref{fig:tumour_subregions} presents a representative contrast-enhanced T1-weighted (T1CE) MRI slice together with its corresponding segmentation mask. The expert-annotated segmentation delineates clinically relevant tumour subregions, including the enhancing tumour, tumour core, and surrounding tumour-associated tissue, which have been shown to capture biologically meaningful characteristics associated with glioma heterogeneity \cite{zeineldin2022multimodal}. These segmented regions guided tumour-aware feature extraction and quantitative radiomics analysis by directing the learning process towards biologically relevant tumour structures. Consequently, the proposed framework captured localised imaging characteristics potentially associated with IDH mutation status and glioma heterogeneity while reducing the influence of surrounding non-tumour brain tissue.

\subsection{Radiomic Feature Extraction}

Radiomic feature extraction was performed to quantify tumour phenotypic characteristics that are not readily captured by deep learning representations alone \cite{maniaci2024integration}. All radiomic features were extracted exclusively from the segmented tumour regions using the expert-annotated tumour masks provided with the BraTS 2021 dataset. Restricting feature extraction to tumour tissue reduces the influence of surrounding healthy brain structures and enables quantitative characterisation of biologically relevant tumour properties \cite{choi2021fully}. Radiomic analysis was conducted using the open-source PyRadiomics library (version 3.0), following the recommendations of the Image Biomarker Standardisation Initiative (IBSI) to ensure feature reproducibility and standardisation \cite{avanzo2020machine,zwanenburg2020image}. Prior to feature extraction, all MRI volumes underwent intensity normalisation and spatial resampling to isotropic voxel spacing. Images were discretised using a fixed bin width to minimise intensity variations across patients while preserving tumour heterogeneity, following standard radiomics preprocessing practices \cite{zwanenburg2020image}. Radiomic features were extracted independently from each of the four MRI modalities (T1, T1CE, T2, and FLAIR). The extracted features comprised multiple complementary categories describing tumour morphology, intensity distribution, and intratumoural heterogeneity, including:

\begin{itemize}
    \item \textbf{First-order statistics}: mean intensity, median, variance, skewness, kurtosis, entropy, energy, and percentile-based intensity measures.
    
    \item \textbf{Shape descriptors}: tumour volume, surface area, compactness, sphericity, elongation, flatness, and maximum diameter.
    
    \item \textbf{Texture features}: Gray-Level Co-occurrence Matrix (GLCM), Gray-Level Run Length Matrix (GLRLM), Gray-Level Size Zone Matrix (GLSZM), Gray-Level Dependence Matrix (GLDM), and Neighbouring Gray Tone Difference Matrix (NGTDM) features, capturing spatial intensity relationships and tumour heterogeneity.
    
    \item \textbf{Wavelet features}: first-order and texture descriptors extracted from wavelet-transformed images to characterise tumour information across multiple spatial frequencies.
\end{itemize}

Following feature extraction, redundant and low-variance features were removed to reduce dimensionality and minimise overfitting. Continuous radiomic variables were subsequently standardised using Z-score normalisation before multimodal feature fusion. The selected radiomic feature vector was concatenated with deep imaging representations and clinical variables to form the final multimodal representation used for prediction of IDH mutation status. Integrating handcrafted radiomic descriptors with deep learning features provides complementary information describing tumour morphology, texture, and spatial heterogeneity that may not be fully represented by convolutional feature extraction alone \cite{maniaci2024integration,choi2021fully}. This multimodal representation enables more comprehensive characterisation of glioma biology and has been shown to improve molecular tumour prediction by exploiting complementary imaging information \cite{boehm2022harnessing,niu2025mri}.

\subsection{Proposed Tumour-Aware Multimodal Framework}

The proposed framework was designed to integrate complementary imaging, radiomic, and clinical information within a unified deep learning architecture for accurate and explainable prediction of IDH mutation status. Unlike conventional MRI-based approaches that analyse the entire brain volume, the proposed framework explicitly focuses feature learning on biologically relevant tumour regions through tumour-aware segmentation, thereby reducing the influence of surrounding healthy tissue and enhancing representation of tumour-specific imaging characteristics \cite{choi2021fully}.

\begin{sidewaysfigure}
\centering
\includegraphics[width=0.95\textheight]{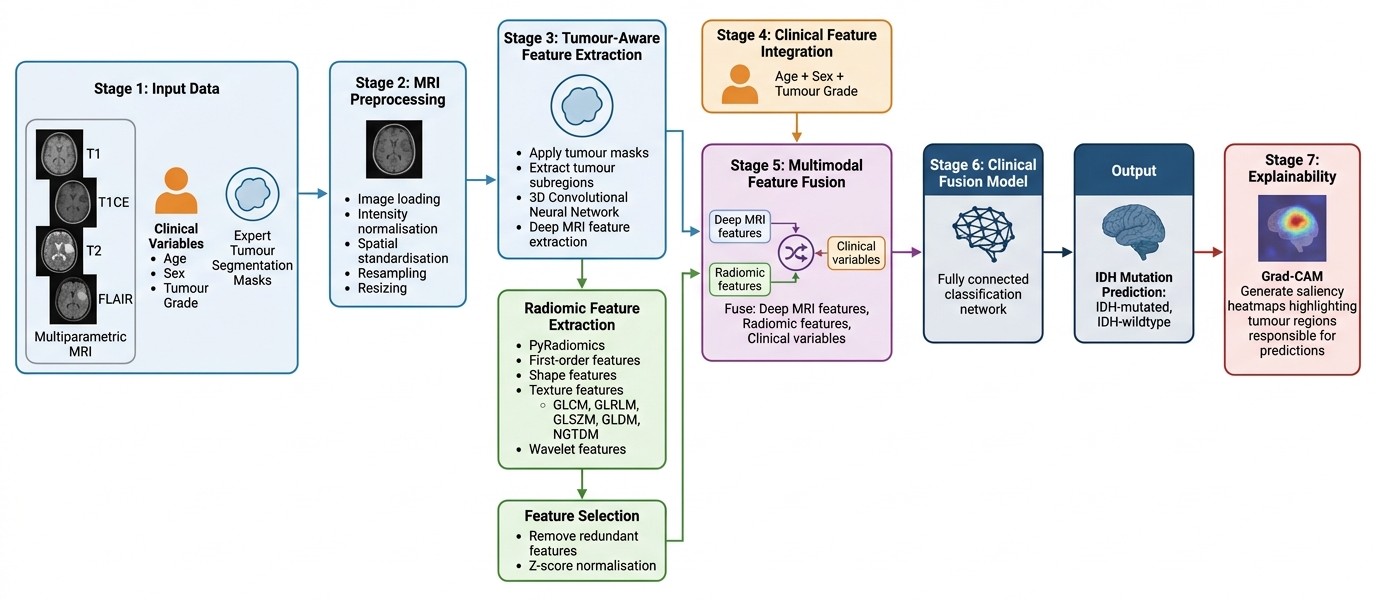}
\caption{Overview of the proposed tumour-aware multimodal framework.}
\label{fig:framework}
\end{sidewaysfigure}

Figure~\ref{fig:framework} illustrates the overall architecture of the proposed tumour-aware multimodal framework. The framework comprises four main components: (i) tumour-aware MRI feature extraction, (ii) radiomic feature extraction, (iii) clinical feature integration, and (iv) multimodal feature fusion and classification. Initially, four MRI modalities (T1, T1CE, T2, and FLAIR), together with their corresponding tumour segmentation masks, are provided as inputs to the imaging branch. Following intensity normalisation and spatial standardisation, tumour masks are applied to guide feature extraction towards biologically relevant tumour subregions. The preprocessed MRI volumes are subsequently processed by a three-dimensional convolutional neural network (3D CNN), which automatically learns hierarchical imaging representations describing tumour morphology, intensity patterns, and structural heterogeneity. In parallel, handcrafted radiomic features are extracted from the segmented tumour regions using the PyRadiomics library \cite{avanzo2020machine}. These features quantitatively characterise tumour phenotype through first-order statistics, shape descriptors, texture matrices, and wavelet-based representations. After feature selection and Z-score normalisation, the resulting radiomic feature vector provides complementary quantitative information describing tumour heterogeneity that may not be fully captured by deep learning features. Clinical information, including patient age, sex, and tumour grade, is incorporated through a dedicated clinical branch. Continuous variables are normalised, whereas categorical variables are encoded prior to integration. 

These variables provide important prognostic information that complements imaging-derived biomarkers and contributes to more comprehensive molecular characterisation. The outputs of the imaging, radiomics, and clinical branches are concatenated to form a unified multimodal feature representation. Fully connected layers subsequently learn interactions among the heterogeneous feature sets before producing the final probability of IDH mutation through a sigmoid activation function. Binary cross-entropy loss is employed to optimise the network parameters during training. To improve transparency and facilitate clinical interpretation, gradient-based class activation mapping (Grad-CAM) is applied after model inference \cite{selvaraju2020grad}. Grad-CAM generates visual heatmaps that highlight the tumour regions contributing most strongly to the predicted molecular status, allowing qualitative assessment of whether the network focuses on biologically meaningful tumour structures rather than irrelevant surrounding tissue. The overall design enables the proposed framework to combine automatically learned imaging representations, handcrafted radiomic biomarkers, and patient-specific clinical characteristics within a single end-to-end prediction pipeline. This multimodal integration is intended to exploit complementary information from multiple sources to improve predictive performance, robustness, and interpretability for non-invasive molecular profiling of gliomas.

\subsection{Experimental Setup}

The proposed tumour-aware multimodal framework was implemented using the PyTorch deep learning framework and trained on a workstation equipped with an NVIDIA GPU supporting CUDA acceleration. Model training was performed using the Adam optimiser with an initial learning rate of 0.0001 \cite{kingma2015adam}. The learning rate was adaptively reduced using a learning rate scheduler when the validation loss plateaued. Binary cross-entropy loss was employed as the optimisation objective for IDH mutation classification. The dataset was randomly divided into training (70\%), validation (15\%), and independent testing (15\%) subsets using stratified sampling to preserve the class distribution of IDH-mutated and IDH-wildtype gliomas across all subsets. Model selection was based exclusively on validation performance, while the independent test set was used only for the final evaluation.

To improve model generalisation and reduce overfitting, several data augmentation techniques were applied during training, including random rotations, horizontal and vertical flipping, intensity perturbation, and random cropping. Early stopping was employed to terminate training when the validation loss did not improve for a predefined number of consecutive epochs. The model parameters corresponding to the lowest validation loss were retained for final testing. Class imbalance between IDH-mutated and IDH-wildtype tumours was addressed using class-weighted binary cross-entropy loss, assigning higher importance to the minority class during optimisation. This strategy reduced prediction bias towards the majority class and improved classification performance for IDH-mutated gliomas.

Hyperparameters were selected empirically based on validation performance. The final model was trained using a batch size of 4 for 100 epochs with a dropout rate of 0.4 applied to the fully connected layers \cite{sabiri2022mechanism}. All experiments were conducted using identical training, validation, and testing partitions to ensure fair comparison between the baseline CNN, tumour-aware ensemble, clinical fusion, and radiomics fusion models. Model performance was evaluated using multiple complementary metrics, including accuracy, precision, recall, F1-score, balanced accuracy, and the area under the receiver operating characteristic curve (AUC). Receiver operating characteristic (ROC) curves, confusion matrices, and ablation studies were further employed to assess model discrimination, classification behaviour, and the individual contribution of each component within the proposed multimodal framework \cite{srivastava2014dropout}. To improve reproducibility, all experiments were performed using a fixed random seed, and identical preprocessing procedures were applied across all evaluated models. The implementation code and trained model weights will be made available upon reasonable request from the corresponding author.

\subsection{Evaluation Metrics}

The predictive performance of the proposed tumour-aware multimodal framework was comprehensively evaluated using several complementary classification metrics. Given the moderate class imbalance between IDH-mutated and IDH-wildtype gliomas, multiple performance measures were employed to provide a balanced assessment of model discrimination rather than relying solely on overall accuracy \cite{sokolova2009systematic}.

The following evaluation metrics were used:

\begin{itemize}
    \item \textbf{Accuracy (ACC):} Measures the proportion of correctly classified samples among all evaluated samples.

    \item \textbf{Precision (PRE):} Measures the proportion of predicted positive samples that are correctly classified as positive.

    \item \textbf{Recall (REC) (Sensitivity):} Measures the proportion of actual positive samples correctly identified by the model.

    \item \textbf{F1-score:} Represents the harmonic mean of precision and recall, providing a balanced assessment of classification performance, particularly for imbalanced datasets.

    \item \textbf{Balanced Accuracy (BAC):} Computes the average of sensitivity and specificity, reducing the influence of class imbalance.

    \item \textbf{Area Under the Receiver Operating Characteristic Curve (AUC):} Measures the model's ability to discriminate between IDH-mutated and IDH-wildtype gliomas across all possible classification thresholds.
\end{itemize}

The evaluation metrics were calculated from the confusion matrix consisting of true positives (TP), true negatives (TN), false positives (FP), and false negatives (FN). Their mathematical definitions are given as follows:

\begin{equation}
ACC=\frac{TP+TN}{TP+TN+FP+FN}
\end{equation}

\begin{equation}
PRE=\frac{TP}{TP+FP}
\end{equation}

\begin{equation}
REC=\frac{TP}{TP+FN}
\end{equation}

\begin{equation}
F1=\frac{2\times PRE \times REC}{PRE+REC}
\end{equation}

\begin{equation}
BAC=\frac{Sensitivity+Specificity}{2}
\end{equation}

The receiver operating characteristic (ROC) curve was generated by plotting the true positive rate against the false positive rate across varying classification thresholds, and the corresponding area under the ROC curve (AUC) was computed to quantify the overall discriminative performance of the model \cite{gonccalves2014roc}. An AUC value of 1.0 indicates perfect discrimination, whereas an AUC of 0.5 corresponds to random classification. To further evaluate the effectiveness of the proposed framework, confusion matrices were analysed to investigate classification errors, while ROC curves were used to compare the discriminative performance of the baseline CNN, tumour-aware ensemble, clinical fusion, and radiomics fusion models. In addition, an ablation study was conducted to quantify the individual contributions of tumour-aware learning, radiomic descriptors, and clinical variables to the overall predictive performance.

\section{Results}

This section presents a comprehensive evaluation of the proposed tumour-aware multimodal framework for non-invasive prediction of IDH mutation status in glioma patients. Model performance was assessed using an independent test set and multiple evaluation metrics, including accuracy, precision, recall, F1-score, balanced accuracy, and the AUC. The experiments were designed to investigate the individual and combined contributions of tumour-aware learning, radiomic features, and clinical information to predictive performance. First, the classification performance of the evaluated models is compared using standard performance metrics. Subsequently, an ablation study is conducted to quantify the contribution of each framework component. ROC analysis and confusion matrix evaluation are then presented to provide further insights into model discrimination and classification behaviour. Finally, explainability analysis is performed using saliency-based visualisation techniques, and the proposed framework is compared with existing state-of-the-art methods reported in the literature.

\subsection{Performance Comparison of Proposed Models}

The predictive performance of the proposed framework was evaluated using multiple complementary classification metrics, including accuracy, precision, recall, F1-score, balanced accuracy, and the area under the receiver operating characteristic curve (AUC). The analysis was designed to assess the individual and combined contributions of tumour-aware learning, radiomic descriptors, and clinical information to the prediction of IDH mutation status. Performance comparisons were conducted across four model configurations, ranging from a baseline MRI-only CNN to progressively more comprehensive multimodal fusion models incorporating tumour-aware imaging, radiomic, and clinical features.

\begin{table}[htbp]
\centering
\caption{Performance comparison of the evaluated models for IDH mutation prediction.}
\label{tab:model_comparison}

\resizebox{\textwidth}{!}{%
\begin{tabular}{lccccc}
\hline
\textbf{Model} & \textbf{Accuracy (\%)} & \textbf{Precision (\%)} & \textbf{Recall (\%)} & \textbf{F1-score (\%)} & \textbf{AUC} \\
\hline
Baseline CNN      & 64.89 & 40.63 & 48.15 & 44.07 & 0.684 \\
No-Grade Ensemble & 80.85 & 62.86 & 81.48 & 70.97 & 0.895 \\
Radiomics Fusion  & 80.85 & 62.16 & 85.19 & 71.88 & 0.917 \\
Clinical Fusion   & 86.17 & 70.59 & 88.89 & 78.69 & 0.944 \\
\hline
\end{tabular}%
}
\end{table}

Table~\ref{tab:model_comparison} summarises the predictive performance of the evaluated models for IDH mutation classification. Classification performance improved consistently as tumour-aware learning, radiomic features, and clinical information were progressively incorporated into the proposed framework. The baseline MRI-only CNN achieved the lowest performance, with an accuracy of 64.89\%, a precision of 40.63\%, a recall of 48.15\%, an F1-score of 44.07\%, and an AUC of 0.684. These results indicate that MRI intensity information alone provides limited discrimination of IDH mutation status. Introducing tumour-aware learning substantially improved predictive performance. The No-Grade Ensemble increased the accuracy to 80.85\%, the F1-score to 70.97\%, and the AUC to 0.895, highlighting the importance of directing feature extraction towards biologically relevant tumour regions. The incorporation of radiomic descriptors further improved model discrimination. 

Although the Radiomics Fusion model achieved the same overall accuracy (80.85\%), its higher recall (85.19\%), F1-score (71.88\%), and AUC (0.917) suggest that radiomic features provide complementary quantitative information beyond deep imaging representations. The Clinical Fusion model achieved the best overall performance, obtaining an accuracy of 86.17\%, a precision of 70.59\%, a recall of 88.89\%, an F1-score of 78.69\%, and an AUC of 0.944. These results indicate that integrating tumour-aware learning, radiomic descriptors, and clinical information yields the strongest predictive performance for non-invasive prediction of IDH mutation status. Overall, the results demonstrate that multimodal feature integration consistently enhanced predictive performance compared with the baseline MRI-only model. The progressive improvements observed across the evaluated configurations suggest that tumour-aware learning, radiomic descriptors, and clinical variables each contribute complementary information, resulting in the highest performance for the proposed Clinical Fusion framework.

\subsection{Ablation Study}

An ablation study was conducted to quantify the individual contributions of tumour-aware learning, radiomic descriptors, and clinical information to the prediction of IDH mutation status. Starting from the baseline MRI-only CNN, additional framework components were progressively incorporated to evaluate their impact on classification performance and discriminative ability. This analysis was designed to assess the relative contribution of each information source and to determine whether multimodal feature integration provides measurable performance improvements over a conventional MRI-only deep learning framework.

\begin{table}[htbp]
\centering
\caption{Ablation configurations of the proposed multimodal framework.}
\label{tab:ablation}

\small
\begin{tabular}{p{3.8cm}p{6.5cm}}
\hline
\textbf{Configuration} & \textbf{Components Included} \\
\hline
Baseline CNN &
MRI only \\

Tumour-aware Ensemble &
MRI + Tumour-aware learning \\

Radiomics Fusion &
MRI + Tumour-aware learning + Radiomic features \\

Clinical Fusion (Proposed) &
MRI + Tumour-aware learning + Radiomic features + Clinical variables \\
\hline
\end{tabular}
\normalsize
\end{table}

Table~\ref{tab:ablation} summarises the progressive integration of tumour-aware learning, radiomic features, and clinical information within the proposed multimodal framework. Starting from the baseline MRI-only CNN, additional components were incorporated sequentially to evaluate their contribution to the prediction of IDH mutation status. Each additional component resulted in improved classification performance. Introducing tumour-aware learning increased the AUC from 0.684 to 0.895, highlighting the value of directing feature extraction towards biologically relevant tumour regions. Incorporating radiomic features further increased the AUC to 0.917, suggesting that quantitative tumour descriptors provide complementary information beyond deep imaging representations. Finally, integrating clinical information achieved the best overall performance, with an accuracy of 86.17\%, an F1-score of 78.69\%, and an AUC of 0.944. Overall, the ablation study indicates that each component contributes incrementally to the predictive performance of the proposed framework. The combined integration of tumour-aware learning, radiomic features, and clinical information provides a more comprehensive representation of glioma characteristics and yields the strongest predictive performance for non-invasive prediction of IDH mutation status.

\subsection{ROC Analysis}

ROC analysis was performed to evaluate the discriminative capability of the evaluated models across different classification thresholds. The area under the receiver operating characteristic curve (AUC) was used to quantify each model's ability to distinguish between IDH-mutated and IDH-wildtype gliomas \cite{gonccalves2014roc}

\begin{figure}[htbp]
\centering
\includegraphics[width=\textwidth]{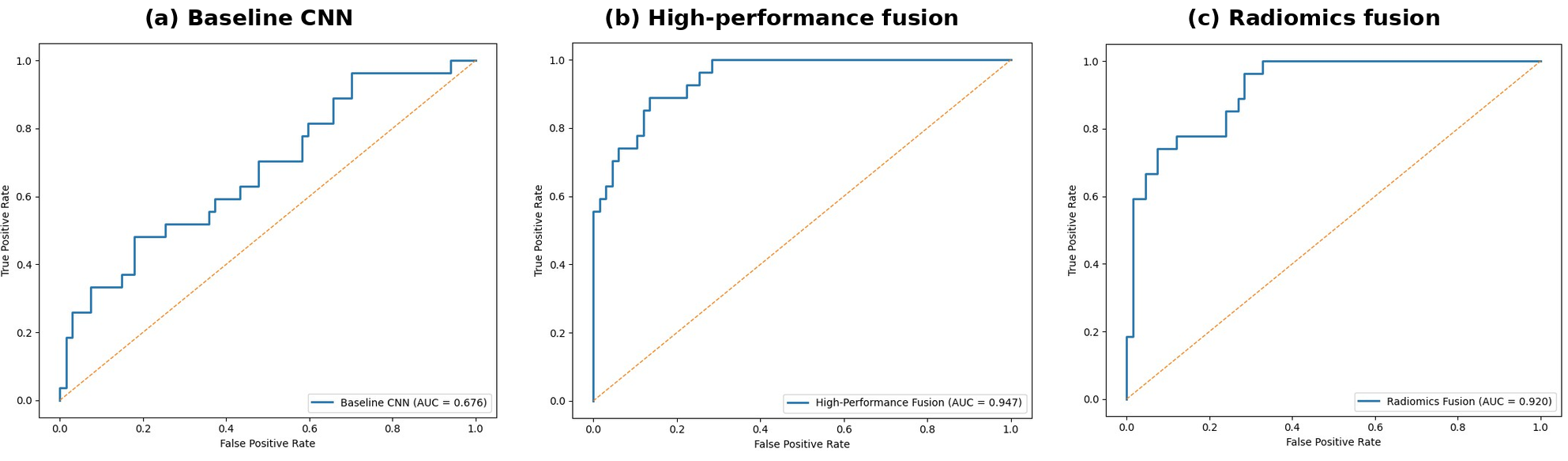}
\caption{ROC curves of the evaluated models for IDH mutation prediction.}
\label{fig:roc}
\end{figure}

Figure~\ref{fig:roc} presents the ROC curves of the evaluated models. The baseline MRI-only CNN achieved the lowest discriminative performance, with an AUC of 0.684, indicating limited ability to distinguish IDH-mutated from IDH-wildtype gliomas using MRI intensity information alone. The No-Grade Ensemble improved the AUC to 0.895, highlighting the benefit of tumour-aware learning for molecular prediction. The Radiomics Fusion model further increased the AUC to 0.917, suggesting that quantitative radiomic descriptors provide complementary discriminative information beyond deep imaging representations. The Clinical Fusion model achieved the highest discriminative performance, with an AUC of 0.944. These results indicate that integrating tumour-aware learning, radiomic features, and clinical variables yields the strongest predictive performance for non-invasive prediction of IDH mutation status.

\subsection{Confusion Matrix Analysis}

To further evaluate the classification performance of the proposed Clinical Fusion model, a confusion matrix was generated using the independent test set. The confusion matrix provides a detailed summary of correctly and incorrectly classified cases for each IDH mutation class, enabling a comprehensive assessment of classification performance and the distribution of prediction errors.

\begin{figure}[htbp]
\centering
\includegraphics[width=0.70\textwidth]{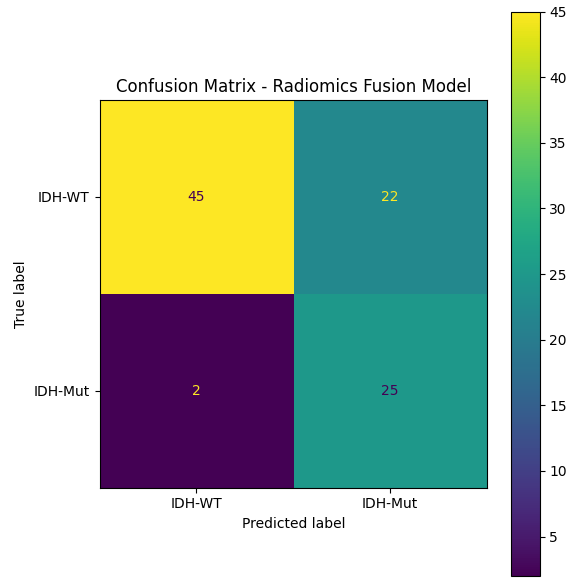}
\caption{Confusion matrix of the proposed Clinical Fusion model}
\label{fig:confusion}
\end{figure}

Figure~\ref{fig:confusion} presents the confusion matrix of the proposed Clinical Fusion model on the independent test set. The model correctly classified 45 IDH-wildtype and 25 IDH-mutated gliomas, while only two IDH-mutated cases were incorrectly classified as IDH-wildtype. The low number of false-negative predictions indicates that the model identified IDH-mutated tumours with high sensitivity, consistent with the high recall achieved by the proposed framework. Most classification errors were false positives, with 22 IDH-wildtype tumours incorrectly classified as IDH-mutated. This pattern suggests that the model tended to assign uncertain cases to the IDH-mutated class, thereby favouring sensitivity over specificity. This behaviour is consistent with the ROC analysis, in which the proposed Clinical Fusion model achieved the highest AUC (0.944). Overall, these results indicate that the proposed Clinical Fusion model provides accurate and interpretable prediction of IDH mutation status.

\subsection{Explainability Analysis}

To investigate the imaging regions that contributed to the model's predictions, gradient-based class activation mapping (Grad-CAM) was performed on the proposed Clinical Fusion model \cite{selvaraju2020grad}. The resulting Grad-CAM heatmaps provide a visual interpretation of the imaging features that contributed most strongly to the final classification decision.

\begin{figure}[htbp]
\centering
\includegraphics[width=\textwidth]{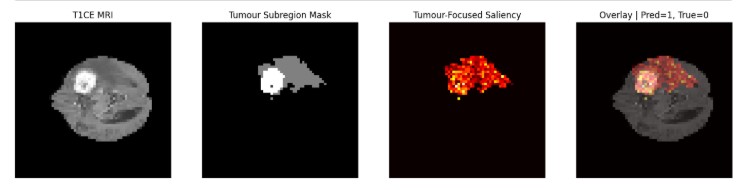}
\caption{Tumour-focused saliency analysis of the proposed Clinical Fusion model.}
\label{fig:saliency}
\end{figure}

Figure~\ref{fig:saliency} presents a representative Grad-CAM visualisation generated by the proposed Clinical Fusion model. The figure highlights the image regions that contributed most strongly to the final prediction. The analysed case corresponded to a correctly classified IDH-wildtype tumour, with a predicted probability of 0.273 for the IDH-mutated class. The Grad-CAM responses were concentrated predominantly within the segmented tumour region, indicating that the model based its prediction primarily on tumour-related imaging characteristics rather than surrounding healthy brain tissue. A strong spatial correspondence was observed between the activated regions and the tumour subregions identified by the segmentation mask. This observation suggests that the proposed framework learned clinically relevant imaging representations associated with tumour morphology and molecular characteristics while minimising reliance on irrelevant background structures. Overall, the Grad-CAM analysis improves model interpretability by providing visual evidence that the proposed framework bases its predictions on biologically meaningful tumour regions.

\section{Discussion}

This study proposed a tumour-aware multimodal framework for non-invasive prediction of IDH mutation status from multiparametric MRI by integrating deep imaging features, quantitative radiomic descriptors, and clinical information. The experimental results demonstrated that the progressive integration of tumour-aware learning, radiomic features, and clinical variables improved predictive performance, with the proposed Clinical Fusion model achieving the highest overall accuracy and discriminative capability. In addition, Grad-CAM analysis improved model interpretability by demonstrating that the network predominantly focused on biologically relevant tumour regions. These findings are discussed in the context of recent advances in MRI-based molecular glioma profiling and existing multimodal deep learning frameworks \cite{christodoulou2025artificial,boehm2022harnessing}. The following sections examine the contribution of the individual framework components, compare the proposed approach with previous studies, discuss its potential clinical implications, and outline the limitations and future research directions.

\subsection{Comparison with Existing Studies}

Table~\ref{tab:comparison} and the experimental results demonstrate that the proposed framework compares favourably with recent MRI-based approaches for preoperative prediction of IDH mutation status. Although direct numerical comparison should be interpreted cautiously because of differences in datasets, patient cohorts, MRI acquisition protocols, preprocessing strategies, and evaluation methodologies, several important methodological observations can be made. Early hybrid approaches demonstrated that combining convolutional neural networks with radiomic features improved molecular prediction compared with deep learning alone. For example, Choi et al. \cite{choi2021fully} proposed a fully automated framework integrating tumour segmentation, CNN-based classification, and radiomic analysis, demonstrating the feasibility of hybrid deep learning–radiomics models for IDH mutation prediction. However, their framework provided limited explainability and did not incorporate patient-specific clinical variables. More recent studies have further expanded multimodal learning through radiomics and transformer-based architectures. Truong et al. \cite{truong2024two} showed that multicontrast MRI radiomics improved preoperative IDH mutation prediction, while Niu et al. \cite{niu2025mri} combined transformer-derived imaging representations with radiomic descriptors to improve molecular classification performance.

Despite these advances, relatively few studies have simultaneously integrated tumour-aware learning, quantitative radiomic descriptors, clinical information, and explainable AI within a unified prediction framework \cite{christodoulou2025artificial,farahani2025diagnostic}. The proposed framework addresses this gap in several important ways. First, feature extraction is explicitly guided by tumour segmentation, reducing the influence of surrounding healthy brain tissue and encouraging biologically meaningful representation learning. Second, the framework integrates complementary information from three independent sources, namely deep imaging features, handcrafted radiomic descriptors, and patient-specific clinical variables. Third, Grad-CAM visualisation provides additional evidence that the model predominantly focused on tumour regions when generating predictions, thereby improving model interpretability.

The proposed Clinical Fusion model achieved an accuracy of 86.17\%, an F1-score of 78.69\%, and an AUC of 0.944, representing the strongest performance among the evaluated models. Although direct comparison with previous studies is inherently challenging because of differences in datasets and evaluation methodologies, the findings suggest that integrating tumour-aware learning, radiomic descriptors, and clinical information provides measurable advantages over approaches relying primarily on imaging features or radiomics alone. Further validation using independent multicentre datasets will be necessary to determine the generalisability of the proposed framework across diverse clinical settings.

\subsection{Impact of Tumour-Aware Learning}

One of the principal findings of this study is the substantial contribution of tumour-aware learning to improving the non-invasive prediction of IDH mutation status. As demonstrated in Table~\ref{tab:ablation}, incorporating tumour-guided feature extraction consistently improved classification performance compared with the baseline MRI-only CNN across all evaluation metrics. By explicitly restricting feature extraction to the segmented tumour region, the proposed framework minimised the influence of surrounding healthy brain tissue and encouraged the network to learn imaging representations that are more closely associated with tumour biology. Compared with the baseline model, the tumour-aware architectures achieved higher accuracy, F1-score, and AUC, indicating improved discrimination between IDH-mutated and IDH-wildtype gliomas. Furthermore, the ROC curves demonstrated a clear improvement in discriminative capability, while the confusion matrix (Figure~\ref{fig:confusion}) showed a reduction in false-negative predictions, suggesting improved sensitivity for identifying IDH-mutated tumours.

These findings are consistent with the growing body of evidence indicating that anatomically guided feature learning can improve the robustness and biological relevance of deep learning models in medical image analysis \cite{choi2021fully,christodoulou2025artificial}. Unlike conventional whole-brain approaches, tumour-aware learning constrains the network to focus on imaging features originating from the lesion itself, thereby reducing interference from normal brain structures. Since tumour regions contain imaging characteristics associated with cellular heterogeneity, tumour morphology, necrosis, oedema, and contrast enhancement, restricting feature extraction to these biologically relevant regions enables the network to learn more discriminative molecular imaging biomarkers \cite{agosti2026mri,rathore2018radiomic}.

The Grad-CAM visualisations shown in Figure~\ref{fig:saliency} further support the effectiveness of the proposed tumour-aware strategy. The activation maps were concentrated predominantly within the segmented tumour regions, suggesting that the model relied primarily on clinically meaningful tumour structures rather than surrounding background tissue when generating predictions. The strong spatial correspondence between the highlighted regions and the tumour masks indicates that the learned representations are biologically plausible and closely associated with imaging characteristics related to IDH mutation status. This level of interpretability is particularly important for clinical translation, as visual explanations of model predictions can improve clinician confidence and facilitate integration of AI-assisted decision support into routine diagnostic workflows \cite{selvaraju2020grad,farahani2025diagnostic}.

Collectively, the ablation study, ROC analysis, confusion matrix evaluation, and Grad-CAM visualisations indicate that tumour-aware learning substantially improved both predictive performance and model interpretability in the proposed framework. Compared with conventional MRI-based deep learning models that rely on global image representations, the proposed tumour-aware framework learned more biologically meaningful imaging features while reducing the influence of irrelevant background tissue. These findings suggest that explicit tumour localisation represents an effective strategy for improving the accuracy, robustness, and clinical interpretability of MRI-based molecular profiling of diffuse gliomas.

\subsection{Contribution of Radiomics and Clinical Features}

The results demonstrate that radiomic features and clinical variables provide important complementary information to the deep imaging features learned by the convolutional network. As shown in Table~\ref{tab:ablation}, incorporating handcrafted radiomic descriptors further improved predictive performance, increasing the AUC from 0.895 to 0.917. This improvement suggests that radiomics captures quantitative imaging characteristics that are not fully represented by deep feature learning alone. Similar observations have been reported in recent radiogenomic studies, where radiomic descriptors were shown to provide complementary information describing tumour morphology, intensity distribution, texture, and intratumoural heterogeneity associated with molecular alterations in gliomas \cite{hosseini2023mri,bera2022predicting,rathore2018radiomic}.

Further improvements were achieved by integrating clinical information, including patient age, sex, and tumour grade. The Clinical Fusion model achieved the highest overall performance, demonstrating that combining tumour-aware learning, radiomic descriptors, and clinical variables provides complementary information for predicting IDH mutation status. Although tumour grade is strongly associated with IDH mutation status in diffuse gliomas, the observed performance improvements cannot be attributed solely to clinical information. The ablation study showed that the tumour-aware model without clinical variables substantially outperformed the baseline MRI-only model, indicating that tumour-aware imaging features contributed independently to prediction performance. The additional improvement achieved by the Clinical Fusion model suggests that clinical variables, including tumour grade, provide complementary prognostic information when integrated with imaging and radiomic features rather than replacing them~\cite{louis20212021,fang2026isocitrate,li2022noninvasive}.

The complementary contribution of these information sources reflects the biological heterogeneity of gliomas. Deep learning effectively captures complex spatial representations directly from MRI, whereas radiomic descriptors provide explicit quantitative measurements of tumour shape, intensity, and texture that characterise tumour heterogeneity \cite{avanzo2020machine,bera2022predicting}. Clinical variables contribute additional patient-level information associated with tumour aggressiveness and disease progression \cite{louis20212021}. Integrating these complementary sources enables the model to learn a more comprehensive representation of tumour biology than can be achieved using any individual modality alone.

The superior performance of the Clinical Fusion model highlights the importance of multimodal integration for molecular prediction. Rather than relying exclusively on imaging-derived features, the proposed framework combines tumour-aware deep learning, handcrafted radiomics, and clinical information within a unified architecture, allowing complementary evidence from multiple domains to contribute to the final prediction. This integrated strategy is consistent with the increasing emphasis on multimodal artificial intelligence in precision oncology, where combining heterogeneous data sources has been shown to improve diagnostic accuracy, robustness, and clinical applicability \cite{boehm2022harnessing,cui2025towards,vanguri2022multimodal}.

Overall, the findings indicate that multimodal feature integration provides a more comprehensive characterisation of gliomas than imaging alone. The observed improvements in predictive performance suggest that combining tumour-aware imaging features, radiomic biomarkers, and routinely available clinical variables represents a promising strategy for non-invasive molecular profiling. Such an approach has the potential to support personalised treatment planning and improve clinical decision-making by providing more accurate preoperative assessment of IDH mutation status.

\subsection{Clinical Relevance of IDH Mutation Prediction}

Accurate non-invasive preoperative prediction of IDH mutation status has considerable clinical importance because IDH is one of the principal molecular biomarkers incorporated into the current World Health Organization (WHO) classification of diffuse gliomas \cite{louis20212021,mcnamara20222021}. Knowledge of IDH mutation status influences tumour classification, prognostic assessment, therapeutic decision-making, and expected patient survival \cite{fang2026isocitrate,li2022noninvasive}. At present, definitive determination of IDH mutation status requires histopathological examination and molecular analysis following surgical biopsy or tumour resection \cite{louis20212021}. Although these approaches remain the clinical gold standard, they are invasive, cannot be repeated frequently during disease monitoring, and are unavailable before treatment planning. Consequently, reliable MRI-based molecular prediction has the potential to provide clinically valuable information before surgery and complement existing diagnostic pathways \cite{khalighi2024artificial,christodoulou2025artificial}.

The proposed tumour-aware multimodal framework demonstrated that integrating deep imaging features, radiomic descriptors, and routinely available clinical variables enables accurate non-invasive prediction of IDH mutation status from standard multiparametric MRI. By combining complementary sources of information, the framework provides a more comprehensive representation of tumour biology than imaging alone. Such an approach may assist clinicians in identifying patients who are more likely to harbour IDH-mutated gliomas, thereby supporting preoperative risk stratification, surgical planning, treatment selection, and multidisciplinary clinical decision-making \cite{boehm2022harnessing,vanguri2022multimodal}.

An important strength of the proposed framework is its interpretability. The Grad-CAM visualisations suggested that model predictions were driven predominantly by tumour tissue rather than surrounding normal brain structures, indicating that the network learned biologically meaningful imaging representations. This level of transparency is particularly important for clinical implementation because explainable AI can improve clinician confidence, facilitate verification of model predictions, and support the responsible integration of AI systems into routine neuro-oncology practice \cite{selvaraju2020grad,abbas2025explainable,holzinger2020measuring}.

The proposed framework is intended to complement rather than replace histopathological examination and molecular testing. Instead, it provides a non-invasive method for estimating IDH mutation status before biopsy or tumour resection, thereby supplying additional information that may assist clinical decision-making. Although routine clinical implementation will require independent external validation, prospective multicentre evaluation, and assessment across diverse imaging protocols, the present findings suggest that tumour-aware multimodal learning represents a promising strategy for supporting precision neuro-oncology. Future integration with clinical decision-support systems and radiology workflows may further enhance personalised diagnosis, treatment planning, and longitudinal management of patients with diffuse gliomas.

\subsection{Limitations and Future Directions}

Several limitations of this study should be acknowledged. First, the proposed framework was developed and evaluated using a single retrospective public dataset. Although the model demonstrated strong predictive performance, its generalisability to independent clinical populations remains to be established. While the BraTS 2021 dataset includes multi-institutional data, model development and evaluation were performed within this single benchmark dataset rather than on an independent external cohort. Variations in MRI acquisition protocols, scanner vendors, imaging parameters, image quality, and patient demographics across institutions may influence model performance. Consequently, external validation using independent prospective multicentre cohorts is essential to assess the robustness, generalisability, and clinical applicability of the proposed framework before routine clinical implementation~\cite{christodoulou2025artificial,farahani2025diagnostic}. Second, the dataset exhibited an imbalanced distribution of IDH-mutated and IDH-wildtype gliomas, reflecting the prevalence observed in clinical practice. Although stratified data partitioning and appropriate optimisation strategies were employed to minimise prediction bias, further evaluation using larger and more balanced datasets would provide additional evidence regarding the robustness and stability of the proposed framework.

Third, the proposed model focused exclusively on predicting IDH mutation status. Contemporary glioma diagnosis increasingly relies on multiple molecular biomarkers, including \textit{1p/19q} codeletion, \textit{MGMT} promoter methylation, and \textit{TERT} promoter mutation, as incorporated into the current WHO classification of central nervous system tumours~\cite{louis20212021,mcnamara20222021}. Extending the proposed framework to simultaneous multi-biomarker prediction would provide a more comprehensive molecular characterisation of diffuse gliomas and increase its clinical utility. Fourth, although the proposed framework incorporated Grad-CAM to improve model interpretability, explainability remains an active area of research. Grad-CAM provides qualitative visual evidence of the image regions contributing to model predictions but does not quantitatively evaluate explanation quality, localisation accuracy, or agreement with expert clinical interpretation. Future studies should investigate complementary explainable AI techniques, such as SHAP, Integrated Gradients, and attention-based visualisation, together with quantitative evaluation methods, including overlap with tumour segmentation masks, localisation metrics, and expert clinician assessment, to provide a more comprehensive evaluation of model interpretability and strengthen confidence in AI-assisted clinical decision support~\cite{selvaraju2020grad,abbas2025explainable,holzinger2020measuring}.

Finally, although the Clinical Fusion model achieved the highest predictive performance, tumour grade was included as one of the clinical variables and is known to be strongly correlated with IDH mutation status~\cite{louis20212021,fang2026isocitrate}. Consequently, the relative contribution of tumour grade to the overall performance improvement cannot be fully distinguished from that of the other imaging-derived and clinical features. Nevertheless, the ablation study demonstrated that tumour-aware learning and multimodal feature integration substantially improved predictive performance before clinical variables were incorporated, suggesting that imaging-derived features made an important contribution to the final model. Future studies should further investigate the relative contributions of tumour grade and other clinical variables through additional ablation experiments, feature importance analyses, and independent external validation. Validation using prospective multicentre cohorts with diverse imaging protocols, scanner vendors, and patient populations will be essential to establish the robustness and generalisability of the proposed framework. In addition, incorporating foundation models for medical imaging, self-supervised representation learning, and transformer-based architectures may further enhance feature representation, robustness, and generalisability~\cite{cui2025towards,zhu2025transformer}. Furthermore, integrating longitudinal imaging, multi-omics data, and additional molecular biomarkers within a unified tumour-aware multimodal framework could facilitate more comprehensive non-invasive molecular profiling of diffuse gliomas, ultimately supporting more accurate diagnosis, personalised treatment planning, and precision neuro-oncology in routine clinical practice.

\section{Conclusion}

This study presented a tumour-aware multimodal framework for explainable, non-invasive prediction of IDH mutation status from multiparametric MRI. By integrating tumour subregion segmentation, quantitative radiomic descriptors, and clinical information within a unified deep learning architecture, the proposed framework addressed important limitations of conventional MRI-based approaches. Comprehensive experimental evaluation demonstrated that tumour-aware learning substantially improved predictive performance, while the integration of radiomic descriptors and clinical variables yielded further gains in classification accuracy and discriminative capability. Grad-CAM visualisations further improved model interpretability by indicating that the network predominantly focused on biologically relevant tumour regions when generating predictions. Collectively, these findings suggest that multimodal feature integration provides a more comprehensive representation of glioma characteristics than MRI intensity information alone, enabling more accurate and reliable non-invasive molecular profiling.

The proposed framework has the potential to complement existing diagnostic pathways by providing preoperative estimation of IDH mutation status to support clinical decision-making and personalised treatment planning. Future research should focus on validating the proposed framework using large independent multicentre cohorts and prospective clinical studies to establish its generalisability and clinical applicability. Extending the framework to predict additional molecular biomarkers and incorporating foundation models, self-supervised representation learning, and transformer-based architectures may further improve predictive performance and facilitate the clinical translation of explainable AI for precision neuro-oncology.

\begin{table}[htbp]
\centering
\caption{List of abbreviations used in this study.}
\label{tab:abbreviations}

\begin{tabular}{ll}
\hline
\textbf{Abbreviation} & \textbf{Definition} \\
\hline
AI & Artificial Intelligence \\
AUC & Area Under the Receiver Operating Characteristic Curve \\
BraTS & Brain Tumor Segmentation Challenge \\
CNN & Convolutional Neural Network \\
FLAIR & Fluid-Attenuated Inversion Recovery \\
F1-score & Harmonic Mean of Precision and Recall \\
GLCM & Gray-Level Co-occurrence Matrix \\
GLDM & Gray-Level Dependence Matrix \\
GLRLM & Gray-Level Run Length Matrix \\
GLSZM & Gray-Level Size Zone Matrix \\
Grad-CAM & Gradient-weighted Class Activation Mapping \\
IDH & Isocitrate Dehydrogenase \\
IBSI & Image Biomarker Standardisation Initiative \\
MGMT & O$^{6}$-Methylguanine-DNA Methyltransferase \\
MRI & Magnetic Resonance Imaging \\
NGTDM & Neighbouring Gray Tone Difference Matrix \\
NIfTI & Neuroimaging Informatics Technology Initiative \\
PyRadiomics & Python Radiomics Library \\
ROC & Receiver Operating Characteristic \\
SHAP & SHapley Additive exPlanations \\
T1 & T1-weighted Magnetic Resonance Imaging \\
T1CE & Contrast-Enhanced T1-weighted Magnetic Resonance Imaging \\
T2 & T2-weighted Magnetic Resonance Imaging \\
TERT & Telomerase Reverse Transcriptase \\
WHO & World Health Organization \\
XAI & Explainable Artificial Intelligence \\
\hline
\end{tabular}
\end{table}

\section*{Acknowledgements}
Not applicable.

\section*{Author Contributions}

Tahsien Al-Quraishi: Conceptualisation, methodology, software, data curation, formal analysis, model development, validation, visualisation, writing (original draft), and writing (review and editing). O. S. Albahri and Raza Nowrozy: Critical review and writing (review and editing). All authors read and approved the final manuscript.

\section*{Funding}
The authors received no specific funding for this work.

\section*{Data Availability}

The BraTS 2021 dataset used in this study is publicly available through the Brain Tumour Segmentation (BraTS) Challenge and can be accessed subject to the dataset usage policies.

\section*{Code Availability}

The implementation code, trained model weights, and supporting scripts developed for this study are available from the corresponding author upon reasonable request for academic research purposes.

\section*{Declarations}

\subsection*{Ethics approval and consent to participate}
Not applicable. This study used publicly available anonymised data and did not involve human participants or animal subjects.

\subsection*{Consent for publication}
Not applicable.

\subsection*{Competing interests}
The authors declare that they have no competing interests.

\end{document}